\documentclass[12pt]{article}

\usepackage{amsmath,amsthm,amssymb,amscd}

\usepackage{graphicx, float, epsfig}
\usepackage[hidelinks]{hyperref}
\usepackage{harvard}
\usepackage{rotating}
\usepackage{ds}

\usepackage{enumitem} 
\usepackage{caption, subcaption} 
\usepackage{capt-of} 
\usepackage[flushleft]{threeparttable} 
\usepackage{multirow}
\usepackage{booktabs}

\usepackage{soul} 
\usepackage[linesnumbered,ruled,vlined]{algorithm2e} 

\usepackage{authblk}

\title{\bf Convex Support Vector Regression}

\author[a]{Zhiqiang Liao}
\author[a,\footnote{Corresponding author. \newline
\hspace*{5mm} \textit{E-mail addresses:} \texttt{zhiqiang.liao@aalto.fi (Z. Liao)}, \texttt{sheng.dai@aalto.fi (S. Dai)},\\
\hspace*{34mm} \texttt{timo.kuosmanen@utu.fi (T. Kuosmanen)}.}]{Sheng Dai}
\author[b]{Timo Kuosmanen}
\affil[a~]{Aalto University School of Business, 02150 Espoo, Finland}
\affil[b~]{Turku School of Economics, University of Turku, 20500 Turku, Finland}
\date{September 2022}

\begin{document}
\captionsetup[figure]{labelfont={bf},labelformat={default},labelsep=period,name={Fig.}}
\captionsetup[table]{labelfont={bf},labelformat={default},labelsep=period,name={Table}}

\citationmode{abbr}
\bibliographystyle{jbes}

\maketitle

\vfill
\vfill

\begin{abstract}
\noindent
Nonparametric regression subject to convexity or concavity constraints is increasingly popular in economics, finance, operations research, machine learning, and statistics. However, the conventional convex regression based on the least squares loss function often suffers from overfitting and outliers. This paper proposes to address these two issues by introducing the convex support vector regression (CSVR) method, which effectively combines the key elements of convex regression and support vector regression. Numerical experiments demonstrate the performance of CSVR in prediction accuracy and robustness that compares favorably with other state-of-the-art methods.
\\[5mm]
\end{abstract}
\vfill

\thispagestyle{empty}
\newpage
\setcounter{page}{1}
\setcounter{footnote}{0}
\pagenumbering{arabic}
\baselineskip 20pt

%

\section{Introduction}\label{sec: intro}

Convex regression (CR) is a classic approach to nonparametric regression that builds upon global concavity or convexity of the regression function (\citename{hildreth1954point}, \citeyear*{hildreth1954point}). Since the explicit piecewise linear characterization of the multivariate model proposed by \citeasnoun{kuosmanen2008representation}, CR has become an active research field with an increasing number of applications in economics, statistics, operational research and related fields (see, e.g., \citename{guntuboyina2018nonparametric}, \citeyear*{guntuboyina2018nonparametric}; \citename{johnson2018shape}, \citeyear*{johnson2018shape}). Recent methodological advances in CR include extensions to quantile-based approaches such as convex quantile regression (\citename{wang2014nonparametric}, \citeyear*{wang2014nonparametric}; \citename{kuosmanen2015stochastic}, \citeyear*{kuosmanen2015stochastic}) and convex expectile regression (\citename{kuosmanen2020how}, \citeyear*{kuosmanen2020how}; \citename{kuosmanen2021shadow}, \citeyear*{kuosmanen2021shadow}). There has been significant development in the computational tools and algorithms (see, e.g., \citename{lee2013a}, \citeyear*{lee2013a}; \citename{mazumder2019a}, \citeyear*{mazumder2019a}; \citename{Dai2022c}, \citeyear*{Dai2022c}; \citename{lin2022augmented}, \citeyear*{lin2022augmented}). 

Overfitting is a longstanding issue in nonparametric methods, including CR. The subgradients fitted by CR can be very large near the boundary of the convex hull of the design points (\citename{seijo2011nonparametric}, \citeyear*{seijo2011nonparametric}; \citename{chen2020on}, \citeyear*{chen2020on}), which can seriously hamper the out-of-sample predictive power. To alleviate overfitting, \cite{lim2014on} has proposed to restrict the domain of the convex hull by imposing additional constraints on the subgradients of the regression function. Another approach bound subgradients is to impose regularization either in objective function or constraints, such as the $L_2$-norm Lipschitz regularization (\citename{mazumder2019a}, \citeyear*{mazumder2019a}) or the $L_\infty$-norm Lipschitz regularization (\citename{balazs2015near}, \citeyear*{balazs2015near}). 

In the literature on machine learning, support vector regression (SVR) is a well-known approach firstly introduced by \citeasnoun{vapnik1999the}. SVR deviates from the linear regression in that it introduces an $\varepsilon$-insensitive loss function instead of the commonly used $L_2$-norm loss function, which helps to improve its out-of-sample performance (\citename{vapnik1999the}, \citeyear*{vapnik1999the}). Therefore, SVR has been considered as a robust alternative against outliers and to reduce overfitting in the context of linear regression.

Thus far, only few studies extend SVR to the context of shape-constrained regression or frontier estimation. The pioneering work by \citeasnoun{wang2012multivariate} was the first one to consider nonparametric convexity-constrained support vector regression (henceforth NCCSVR). In this approach, the Hessian matrix of a nonparametric representor function is constrained to be positive semidefinite in each observation. The authors transform the shape-constrained SVR into a semidefinite programming problem, assuming the regression function to be continuous and twice differentiable throughout its domain. One notable limitation of NCCSVR is that it is not applicable for the univariate CR with a single regressor.

Recently, \citeasnoun{valero-carreras2021support} and \cite{valero2022multi} relax continuity and convexity assumptions, adapting SVR to the nonparametric estimation of frontier production and cost functions that envelop all observations. A notable limitation of their approach is the deterministic nature of the data generating process, which assumes away any noise in data. This observation motivates us to combine the key elements of both CR and SVR in a unified framework.

The main objective of the present paper is to improve the out-of-sample predictive power of CR by alleviating overfitting. To this end, we combine the key characteristics of both CR and SVR in a new approach referred to as convex support vector regression (CSVR). A notable difference between NCCSVR by \citeasnoun{wang2012multivariate} and the proposed CSVR approach concerns the implementation of the convexity constraints: NCCSVR imposes constraints on the Hessian matrix, whereas CSVR makes use of the inequality constraints known as the Afriat inequalities (see \citename{kuosmanen2008representation}, \citeyear*{kuosmanen2008representation}). Using Monte Carlo simulations and two real-world examples, we show that the proposed CSVR approach yields a smaller mean squared error (MSE) than other state-of-the-art methods, including the NCCSVR method.

Our secondary objective is to outline how the proposed CSVR approach can be extended to facilitate the automatic variable selection in applications with high dimensionality. Inspired by such works as \cite{bradley1998feature}, \cite{zhao2009composite} and \cite{negahban2011simultaneous}, two alternative formulations of LASSO CSVR are considered (LASSO refers to the least absolute shrinkage and selection operator). This extension further enhances the linkages between SVR and LASSO that originated in the machine learning literature and CR that has emerged in econometrics and statistics. 

The rest of this paper is organized as follows. Section \ref{sec: prelim} briefly reviews classical statistics and machine learning methods for regression problems. We then introduce the new shape-constrained SVR method, extend it to the Lasso version, and present a graphical illustration in section \ref{sec: csvr}. Section \ref{sec: sim} presents some evidence from Monte Carlo simulations. In the section \ref{sec: app} we experimentally compare CSVR against competing methods on two real-world datasets. Section \ref{sec: con} presents our concluding remarks.

%

\section{Preliminaries on regression}\label{sec: prelim}

\subsection{Convex regression}
Considering a general nonparametric regression model with a set of observations $\{(\mathbf{x}_i, y_i)\}_{i=1}^n$ satisfying
\begin{equation}\label{nonpara}
    y_i = f(\mathbf{x}_i) + \varepsilon_i, \quad \mbox{ for } i = 1, \ldots, n,
\end{equation}
where $\mathbf{x} \in \mathbb{R}^d$ is an observed vector of predictors, $y_i \in \mathbb{R}$ is the response variable, and $\varepsilon_i$ is a random noise with zero mean. The regression function $f: \mathbb{R}^d \rightarrow \mathbb{R}$ in Eq.~\eqref{nonpara} is unknown but satisfies certain shape restrictions such as monotonicity, concavity, and homogeneity (see, e.g., \citename{kuosmanen2010data}, \citeyear*{kuosmanen2010data}; \citename{yagi2020shape}, \citeyear*{yagi2020shape}). In this paper we focus exclusively on the class $\mathcal{F}$ of concave function $f$, that is
\begin{equation*}
 \mathcal{F} := \Big\{f: \mathbb{R}^d \rightarrow \mathbb{R} \mid \forall \mathbf{x}_1, \mathbf{x}_2 \in \mathbb{R}^d, \tau f(\mathbf{x}_1) + (1-\tau) f(\mathbf{x}_2) \le f(\tau \mathbf{x}_1 + (1-\tau) \mathbf{x}_2) \Big\}. 
\end{equation*}

The basic idea of CR is to find the best fitting function $f$ from a family of continuous and concave functions $\mathcal{F}$ by minimizing the sum of squares of the residuals 
\begin{alignat}{2}
	\min  \, &\frac{1}{2}\sum_{i=1}^n(y_i - f(\mathbf{x}_i))^2 \label{eq:cr_1} \\
	\mbox{\textit{s.t.}}\quad 
	&  f \in \mathcal{F} \notag
\end{alignat}

While problem \eqref{eq:cr_1} is the infinite-dimensional multivariate convex regression problem, it can be equivalently represented by a finite-dimensional quadratic programming (QP) problem. Following \citeasnoun{kuosmanen2008representation}, we consider the following least squares estimator as the operational multivariate convex regression model 
\begin{alignat}{2}
    \min_{\boldsymbol{\beta}, \boldsymbol{\alpha}, \boldsymbol{\varepsilon}} \quad & \frac{1}{2} \sum_{i=1}^{n} \varepsilon_i^2  &{}& \label{cr} \\
    \mbox{\textit{s.t.}}\quad
    & y_i = \alpha_i + \boldsymbol{\beta}_i^{'} \mathbf{x}_i + \varepsilon_i &\quad& \forall i \notag \\
    &\alpha_i + \boldsymbol{\beta}_i^{'} \mathbf{x}_i \leq \alpha_h + \boldsymbol{\beta}_h^{'} \mathbf{x}_i &{}& \forall i,h \notag 
\end{alignat}
where $\boldsymbol{\beta}^{'}$ indicates the transpose of $\boldsymbol{\beta}$, and its subscript $h$ is any index of data point not equal to $i$. The first constraint of \eqref{cr} simply restates the regression equation \eqref{nonpara} in terms of a piecewise linear approximation of the true but unknown regression function $f$, and the second constraint enforces concavity of the piecewise linear regression function (reversing the sign of the inequality imposes convexity). Note that additional monotonicity constraints could be implemented by restricting the sign of $\boldsymbol{\beta}$ (e.g., $\boldsymbol{\beta} \geq 0$ for monotonic increasing and $\boldsymbol{\beta} \leq 0$ for monotonic decreasing functions). Further, imposing $\alpha=0$ imposes linear homogeneity (constant returns to scale). See \citeasnoun{kuosmanen2015stochastic} for more detailed discussion.

Given the optimal solutions ($\hat{\alpha}_i$, $\boldsymbol{\hat{\beta}}_i$) to problem \eqref{cr}, we can reconstruct the explicit representor function $\hat{f}^{CR}(\mathbf{x})$ as (\citename{kuosmanen2008representation}, \citeyear*{kuosmanen2008representation})
\begin{equation}\label{rep}
    \hat{f}^{CR}(\mathbf{x})=\min_{i = 1, \ldots, n} \big\{\hat{\alpha}_i+\boldsymbol{\hat{\beta}}_i^{'}\mathbf{x} \big\}
\end{equation}

However, the estimated coefficients $\hat{\boldsymbol{\beta}}_i$ could be arbitrarily large, particularly near the boundary of the convex hull of the covariate domain, due to the fact that the feasible set of problem \eqref{cr} can be unbounded (see, e.g., \citename{mazumder2019a}, \citeyear*{mazumder2019a}; \citename{chen2020on}, \citeyear*{chen2020on}). This may lead to potential overfitting and deteriorate the out-of-sample performance of the estimated function. Even for the univariate case, the estimated $\hat{\beta}_i$ are also unbounded at the boundary (\citename{ghosal2017univariate}, \citeyear*{ghosal2017univariate}).

To alleviate the overfitting problem in convex regression \eqref{cr}, \cite{mazumder2019a} apply the $L_2$-norm Lipschitz regularization on subgradients with a known bound $L>0$ to reduce overfitting. For a prespecified $L>0$, the class of $F_L$ of concave functions with Lipschitz regularization
\begin{equation*}
    F_L:= \{f: \mathbb{R}^d \rightarrow \mathbb{R} | f \text{ } \mbox{is concave}; \underset{\mathbf{x} \in \mathbb{R}^d}{\sup} \| \partial f(\mathbf{x})\| \le L \}.
\end{equation*}
where $\| \partial f(\mathbf{x}) \|$ is the maximum of $\| \cdot \|_2$-norm of vectors in $\partial f(\mathbf{x})$. The corresponding Lipschitz convex regression (LCR) is then formulated as 
\begin{alignat}{2}
    \min_{\boldsymbol{\beta}, \boldsymbol{\alpha}, \boldsymbol{\varepsilon}} \quad &  \frac{1}{2} \sum_{i=1}^{n} \varepsilon_i^2 &{}& \label{lcr} \\
    \mbox{\textit{s.t.}}\quad
    & y_i = \alpha_i + \boldsymbol{\beta}_i^{'} \mathbf{x}_i + \varepsilon_i  &\quad& \forall i \notag \\
    &\alpha_i + \boldsymbol{\beta}_i^{'} \mathbf{x}_i \leq \alpha_h + \boldsymbol{\beta}_h^{'} \mathbf{x}_i &{}& \forall i,h \notag  \\
    &\|\boldsymbol{\beta}_i\|_2 \leq L &{}&  \forall i  \notag 
\end{alignat}

Note that the estimated function $\hat{f}^{LCR}$ can be obtained by inserting the optimal $\hat{\alpha}_i$ and $\boldsymbol{\hat{\beta}}_i$ from \eqref{lcr} to the equation of explicit representor function \eqref{rep}. We can also resort to other Lipschitz norms (e.g.,  $\| \cdot \|_\infty$-norm) to address the overfitting problem in convex regression (\citename{lim2014on}, \citeyear*{lim2014on}; \citename{balazs2015near}, \citeyear*{balazs2015near}). Following \cite{mazumder2019a}, $\| \cdot \|_2$-norm is more effective than $\| \cdot \|_\infty$-norm in avoiding overfitting. However, when the data with high noise, the outliers and the large variance in the error term could weaken the effectiveness of the Lipschitz norm approaches in reducing overfitting. Such a gap also motivates this paper to propose a new convex support vector regression approach and empirically compare the proposed method with other related approaches in terms of finite sample performance.

\subsection{Support vector regression}\label{subsec: svr}

Support vector regression (SVR) belongs to the class of support vector machines. As a regression method, it aims to estimate a function $f(x)$ that follows the structural risk minimization principle grounded on the statistical learning theory. It gives good generalization capacity by minimizing the upper bound of the risk (\citename{vapnik1999the}, \citeyear*{vapnik1999the}).

SVR has excellent potential to reduce overfitting because the structural risk minimization principle makes a trade-off between the prediction accuracy and the complexity of the regression function. When some unwanted data exit, some parameters in the conventional regression models may become large to accommodate such outliers. Thus, these models may suffer from overfitting, while the SVR has good generalization performance as its optimization object function guarantees the flatness of the regression function. Flatness in the regression model means that one minimizes vector $\boldsymbol{\beta}$. Taking into account data errors, one can introduce the slack variables $\xi$ and $\xi^*$ and penalty term $C$. The slack variables can be used to allow some errors that lie on the outside of the margin $\varepsilon$ which refers to $\textit{soft margin}$ (\citename{cortes1995support}, \citeyear*{cortes1995support}). Hence we can obtain the regression function by solving the following optimization problem
\begin{alignat}{2}
    \min_{\boldsymbol{\beta}, \alpha, \xi_i, \xi_i^*} \quad &\frac{1}{2}\|\boldsymbol{\beta} \|^2_2 + C\sum_{i=1}^{n} (\xi_i + \xi_i^*) &{\quad}& \label{svr} \\
    \mbox{\textit{s.t.}}\quad
    & y_i - \alpha - \boldsymbol{\beta}^{'} \mathbf{x}_i \leq \varepsilon + \xi_i  &{}&  \forall i  \notag  \\
    &\alpha + \boldsymbol{\beta}^{'} \mathbf{x}_i - y_i \leq \varepsilon + \xi_i^* &{}&  \forall i  \notag  \\
    &\xi_i \geq 0 , \xi_i^* \geq 0 &{}&  \forall i  \notag 
\end{alignat}
where $C$ is a prespecified parameter that determines the trade-off between the complexity of regression function $f(x)$ and the prediction accuracy $\varepsilon$. This corresponds to the most commonly adopted $\varepsilon$-insensitive loss function $\lvert\xi\rvert_{\varepsilon}$ described by
\begin{equation}\label{loss}
    |\xi|_{\varepsilon}= \begin{cases} 
    0, & \text{if $|\xi| \leq \varepsilon$;} \\
    |\xi|-\varepsilon, & \text{otherwise.}
    \end{cases}
\end{equation}
for a user-determined nonnegative number $\varepsilon$. A potential benefit of using $\varepsilon$-insensitive loss function is robustness to outliers because it is less sensitive to noisy inputs. In practice, the penalty constant $C$ and parameter $\varepsilon$ can be chosen based on the user's experience. It can also be determined by cross-validation, a standard model selection technique in machine learning. Furthermore, the estimated support vector function is simply $\hat{f}^{SVR}(\mathbf{x}) = \alpha + \boldsymbol{\beta}^{'}\mathbf{x}$.

\section{Combining CR and SVR}\label{sec: csvr}

\subsection{Convex support vector regression (CSVR)}

To address the overfitting problem and improve the model robustness, we blend the key elements of CR and SVR and propose the convex support vector regression (CSVR) approach. Consider the following quadratic programming problem

\vspace{-2em}
\begin{alignat}{2}
    \min_{\boldsymbol{\beta}_i, \alpha, \xi, \xi^*} \quad &\frac{1}{2}\sum_{i=1}^{n}\|\boldsymbol{\beta}_i\|_2^2 + C\sum_{i=1}^{n} (\xi_i + \xi_i^*) &{\quad}& \label{csvr} \\
    \mbox{\textit{s.t.}}\quad
    & y_i - \alpha_i - \boldsymbol{\beta}_i^{'} \mathbf{x}_i \leq \varepsilon + \xi_i  &{}&  \forall i  \notag \\
    &\alpha_i + \boldsymbol{\beta}_i^{'} \mathbf{x}_i - y_i \leq \varepsilon + \xi_i^* &{}&  \forall i  \notag \\
    &\alpha_i + \boldsymbol{\beta}_i^{'} \mathbf{x}_i \leq \alpha_h + \boldsymbol{\beta}_h^{'} \mathbf{x}_i &{}&  \forall i, h  \notag \\
    &\xi_i \geq 0 , \xi_i^* \geq 0 &{}&  \forall i  \notag 
\end{alignat}
where the first two constraints restrict the error terms into a specified margin, noted as the maximum error ($\varepsilon$), and consider the possible outliers with the deviation from the margin as $\xi_i$. The third constraint, a system of Afriat inequalities, guarantees the concavity of the unknown function $f$. Compared to problem (\ref{svr}), problem (\ref{csvr}) is a shape constrained extension by means of Afriat inequalities. Note that the coefficients $\boldsymbol{\beta}_i$ represent the subgradient of the concave function $f$ at point $\mathbf{x}_i$. 

To further investigate the relationship between the CSVR and regularized function estimation, we rewrite problem \eqref{csvr} as the following equivalent formulation
\begin{alignat}{2}
    \min_{\boldsymbol{\beta}_i, \alpha} \quad & \sum_{i=1}^{n} |y_i - \alpha_i - \boldsymbol{\beta}_i^{'} \mathbf{x}_i|_{\varepsilon} + \frac{A}{2}\sum_{i=1}^{n}\|\boldsymbol{\beta}_i\|^2_2 
    \label{csvre}\\
    \mbox{\textit{s.t.}}\quad
    & \alpha_i + \boldsymbol{\beta}_i^{'} \mathbf{x}_i \leq \alpha_h + \boldsymbol{\beta}_h^{'} \mathbf{x}_i \quad \forall i, h \notag
\end{alignat}
where $A$ is the tuning parameter, playing a similar role as $C$. The function $|\cdot|_{\varepsilon}$ indicates the $\varepsilon$-insensitive loss function described by equation \eqref{loss}. Note that the objective function has a form of \textit{loss}+\textit{penalty}; hence, the parameter $A$ controls the trade-off between loss and penalty. The standard cross-validation techniques can be used to determine the tuning parameter. 

The penalty term in problem \eqref{csvre} is a $L_2$-norm of the subgradient vector, the same as the penalized convex regression using the norm of the subgradients (\citename{aybat2014a}, \citeyear*{aybat2014a}; \citename{lim2014on}, \citeyear*{lim2014on}; \citename{Dai2022}, \citeyear*{Dai2022}). The ridge penalty shrinks the subgradients towards zero, which means the regression function will be as flat as possible. This shrinkage can control the variance of the subgradients, thus helping to alleviate the overfitting problem via the bias-variance trade-off, especially when many highly correlated variables exist. We will further study the effect of regularization in the following section.

Using the $\varepsilon$-insensitive loss function also enables a sparse set of support vectors to be obtained. That is, the number of support vectors increases more slowly than linearly. As an extended SVR model, the CSVR approach retains this advantage of sparsity. Moreover, the $\varepsilon$-insensitive loss function is less sensitive to outliers than the quadratic loss function used in problem \eqref{lcr}. CSVR is more robust than conventional CR when there are outliers in the dataset, therefore, leading to a smaller prediction error than the usual CR; see Section \ref{sec: sim} for simulation evidence. When $\varepsilon=0$, we can obtain a special case of $L_1$ loss function in problem \eqref{csvre}. Note that the $L_1$ loss function is relatively more robust to outliers than the quadratic loss function (\citename{alquier2019estimation}, \citeyear*{alquier2019estimation}). 

To derive a similar version to problem \eqref{lcr}, letting $\phi_{\varepsilon}(\cdot)$ be the $\varepsilon$-insensitive loss function, problem \eqref{csvre} can be rewritten as fully constrained optimization problem as follows
\begin{alignat}{2}
    \min_{\boldsymbol{\beta}_i, \alpha} \quad & \sum_{i=1}^{n} \phi_{\varepsilon}(y_i - f(\mathbf{x}_i)) \label{csvre2}\\
    \mbox{\textit{s.t.}}\quad
    & \alpha_i + \boldsymbol{\beta}_i^{'} \mathbf{x}_i \leq \alpha_h + \boldsymbol{\beta}_h^{'} \mathbf{x}_i \quad \forall i, h \notag\\
    & \sum_{i=1}^{n}\|\boldsymbol{\beta}_i\|_2 \leq C \notag
\end{alignat}
where $\sum_{i=1}^{n}\|\boldsymbol{\beta}_i\|_2 \leq C$ is a $L_2$-norm penalty, and $C$ is a nonnegative tuning parameter. We note here that the $L_2$-norm penalty used in this study differs from the $L_2$-norm Lipschitz regularization in \cite{mazumder2019a}. \cite{mazumder2019a} consider a least square estimator over a set of convex functions that are uniformly Lipschitz with a known bound (see problem \eqref{lcr}), whereas problem \eqref{csvre2} is ridge regularized. Also, note that we use a different loss function, whose insensitivity may contribute to the performance of CSVR. 

\subsection{Lasso CSVR}

In addition to $L_2$-norm, we could introduce other regularization methods such as $L_1$-norm to extend the present CSVR approach. The $L_1$-norm support vector machine was first proposed by \cite{bradley1998feature} for solving classification problems. Inspired by this idea, we briefly describe an extension of CSVR: the Lasso CSVR model for the variable selection regression analysis. This extension provides an alternative path of extending CSVR to select variables automatically. The first version of Lasso CSVR replaces the $L_2$-norm penalty in problem \eqref{csvre} with $L_1$-norm penalty
\begin{alignat}{2}
    \min_{\boldsymbol{\beta}_i, \alpha} \quad & \sum_{i=1}^{n} |y_i - \alpha_i - \boldsymbol{\beta}_i^{'} \mathbf{x}_i|_{\varepsilon} + \frac{A}{2}\sum_{i=1}^{n}\|\boldsymbol{\beta}_i\|_1 
    \label{csvrl1}\\
    \mbox{\textit{s.t.}}\quad
    & \alpha_i + \boldsymbol{\beta}_i^{'} \mathbf{x}_i \leq \alpha_h + \boldsymbol{\beta}_h^{'} \mathbf{x}_i \quad \forall i, h \notag
\end{alignat}
Similar to the $L_2$-norm penalty, the $L_1$-norm penalty can control the variance of the estimation and improve prediction accuracy. Moreover, the Lasso performs automatic variable selection, which is not the case for the $L_2$-norm penalty. Although the performance of Lasso does not uniformly dominate the ridge regression (\citename{tibshirani1996regression}, \citeyear*{tibshirani1996regression}), the $L_1$-norm CSVR appears very promising because the variable selection is increasingly important in modern data science.

The variable selection aspect of the $L_1$-norm CSVR approach is useful for regression analysis in the case that there exist no highly correlated variables (see the limitations of $L_1$-norm penalty in \citename{zou2005regularization}, \citeyear*{zou2005regularization}). While the $L_1$-norm penalty can shrink the subgradients of the function and make the subgradients of irrelevant variables small, it cannot reduce them to zero exactly. The reason is that selecting variables by regularizing the subgradient $\boldsymbol{\beta}_i$ in problem \eqref{csvrl1} with a group sparsity penalty is not an effective way due to the existing of Afriat inequalities (\citename{xu2016faithful}, \citeyear*{xu2016faithful}; \citename{Dai2022c}, \citeyear*{Dai2022c}). That is, the small changes to $\boldsymbol{\beta}_i$ in each Afrait inequality $\alpha_i + \boldsymbol{\beta}_i^{'} \mathbf{x}_i \leq \alpha_h + \boldsymbol{\beta}_h^{'} \mathbf{x}_i$ may not violate the concavity assumption. Therefore, the $L_1$-norm CSVR approach can make certain coefficients very small but not be zero, and thus does not necessarily make the representor function \eqref{rep} sparse.

This motivates us to consider a $L_\infty$-regularized Lasso CSVR. As shown in \cite{zhao2009composite} and \cite{negahban2011simultaneous}, the $L_\infty$-norm taking the maximum encourages all $d$ components of the subgradient $\boldsymbol{\beta}_i$ to be zero simultaneously or to be
nonzero simultaneously. The second version of Lasso CSVR can be formulated as the following optimization problem
\begin{alignat}{2}
    \min_{\boldsymbol{\beta}_i, \alpha} \quad & \sum_{i=1}^{n} |y_i - \alpha_i - \boldsymbol{\beta}_i^{'} \mathbf{x}_i|_{\varepsilon} + \frac{A}{2}\sum_{i=1}^{n}\|\boldsymbol{\beta}_i\|_\infty 
    \label{csvrinf}\\
    \mbox{\textit{s.t.}}\quad
    & \alpha_i + \boldsymbol{\beta}_i^{'} \mathbf{x}_i \leq \alpha_h + \boldsymbol{\beta}_h^{'} \mathbf{x}_i \quad \forall i, h \notag
\end{alignat}

Compared to the $L_1$-norm penalty, the $L_\infty$-norm penalty restricts each subgradient $\beta_{j,i}$ with a fixed bound instead of all $d$ components of the subgradient $\boldsymbol{\beta}_i$. Furthermore, the $L_\infty$-regularized Lasso approach is a special case of the block $L_1 / L_\infty$-regularization when the number of blocks is $n$ in \cite{negahban2011simultaneous}.

\subsection{Illustrative example}

We proceed to demonstrate how the fitted functions estimated by CSVR and convex regression are different and illustrate the potential advantages of CSVR in reducing overfitting via an artificial example. In so doing, we generate 50 observations with $y_i=3+\ln(x_i)+\epsilon_i$, where $x_i$ is randomly drawn from $U(1,10)$ and the error term $\epsilon_i$ is generated independently from $N(0, 0.7^2)$. To search the optimal hyperparameters, we resort to the standard fivefold cross-validation approach (see, e.g., \citename{mazumder2019a}, \citeyear*{mazumder2019a}; \citename{Dai2022c}, \citeyear*{Dai2022c}).

\begin{figure}[H]
    \centering
    \includegraphics[width=0.6\textwidth]{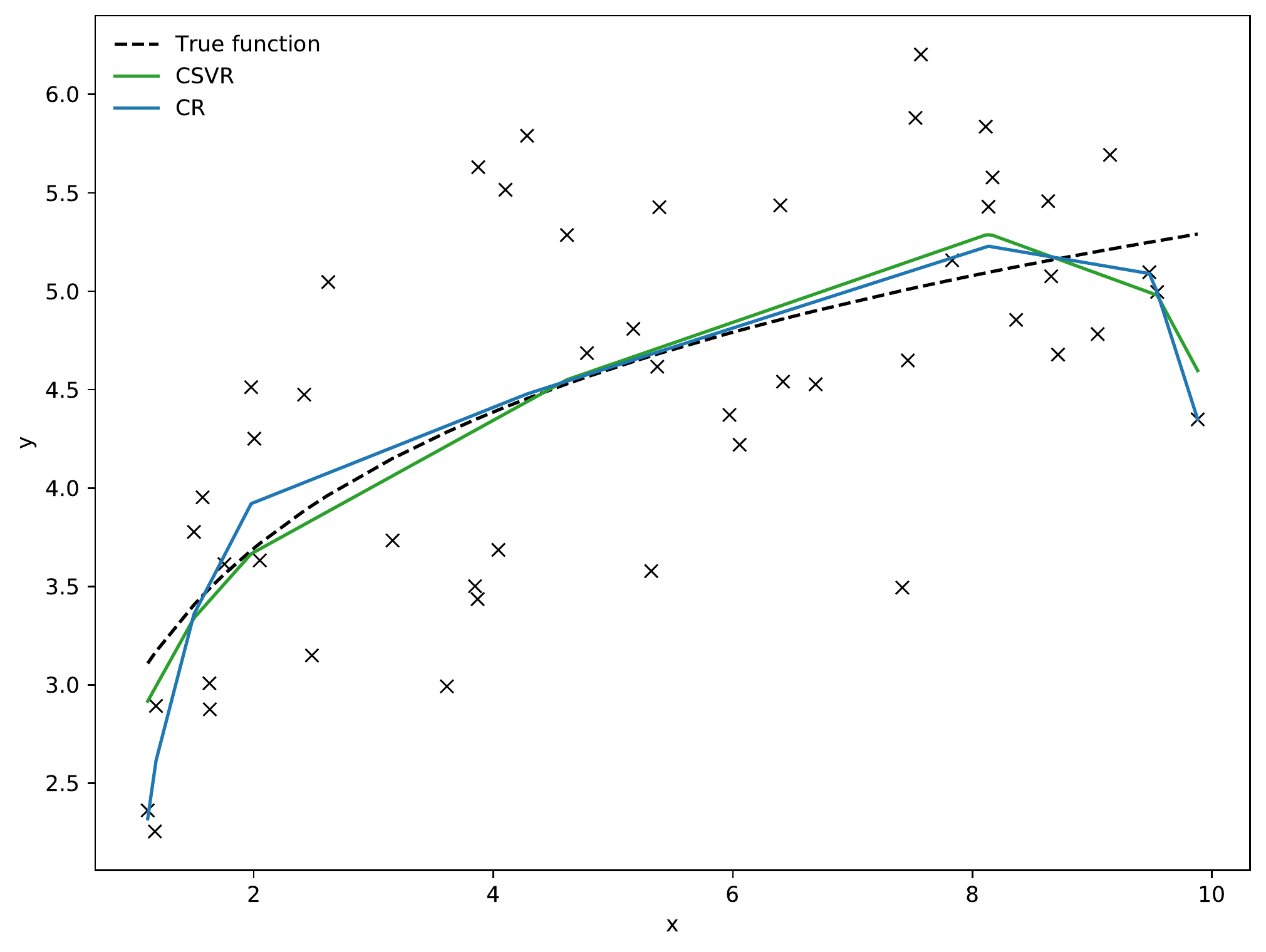}
    \caption{Illustration of the fitted functions estimated by CSVR and convex regression.}
    \label{fig: cov}
\end{figure}

Fig.~\ref{fig: cov} depicts the fitted functions estimated by CSVR and convex regression. We observe that both shape-constrained approaches yield piecewise concave lines and can capture the shape of data points. The fitted convex regression function appears to be more sharper, while the fitted CSVR functions seem to be relatively smooth, implying the difference in the loss function. When comparing the approximation errors, CR has a smaller in-sample error (i.e., $MSE=0.44$) in comparison with CSVR. However, a smaller training error may lead to a larger test error, suggesting that CR is hampered by overfitting, and CSVR can effectively avoid this problem.

\begin{figure}[H]
    \centering
    \includegraphics[width=0.6\textwidth]{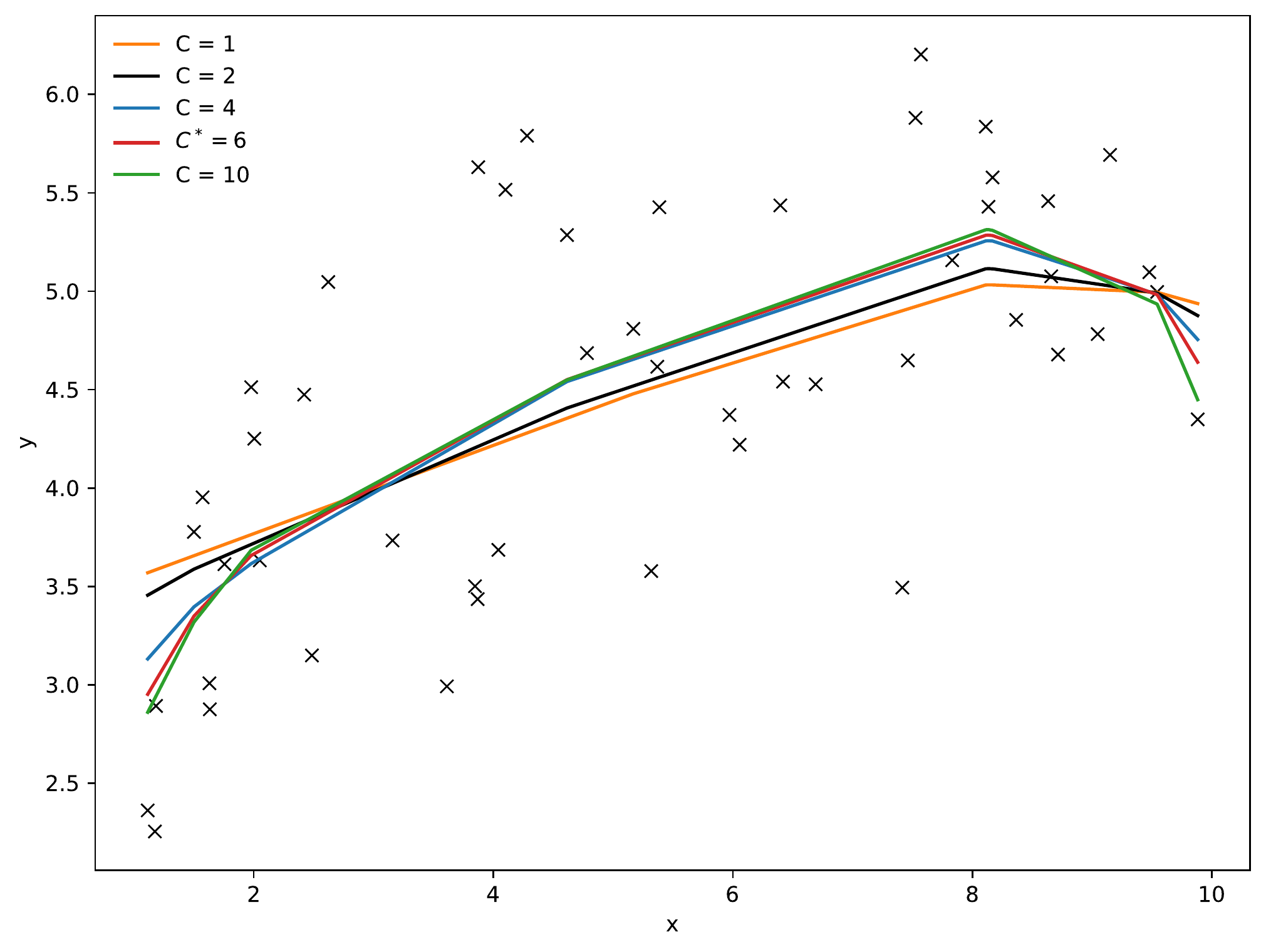}
    \caption{Illustration of the estimated CSVR functions with different values of $C$.}
    \label{fig: para}
\end{figure}

We then illustrate the robustness of the estimated CSVR function to the choice of the tuning parameter $C$. In Fig.~\ref{fig: para}, the hyperparameter $\varepsilon$ is fixed at 0.1 and $C$ is tuned over 5 values from the set $\{1, 2, 4, 6, 10\}$. Note that the optimal hyperparameter $C^*=6$ is determined by the standard cross-validation approach. It is evidently from Fig.~\ref{fig: para} that the parameter $C$ can reshape the estimated CSVR functions but produce very similar estimated piecewise-linear curves.

%

\section{Monte Carlo study}\label{sec: sim}

Having illustrated the estimated CSVR function, we proceed to investigate the finite sample performance of the CSVR, CR, SVR with radial basis function kernel, and LCR approaches in the controlled environment of Monte Carlo simulations. The main objective of our simulations is to examine whether the proposed CSVR approach can better fit the true function by addressing the overfitting problem.

\subsection{Setup}

Consider the following data generating processes (DGP) (see, e.g., \citename{valero-carreras2021support}, \citeyear*{valero-carreras2021support})
\begin{enumerate}[label= \arabic*), leftmargin=1.2cm]
    \item DGP I: \(y = 3+x_1^{0.5}+\epsilon\)
    \item DGP II: \(y = 3+x_1^{0.2}+x_2^{0.3}+\epsilon\)
    \item DGP III: \(y = 3+x_1^{0.05}+x_2^{0.15}+x_3^{0.3}+\epsilon\)
\end{enumerate}
where $x_1$, $x_2$, and $x_3$ are independently and randomly sampled from the uniform distribution $U[1,10]$ and the error term $\epsilon$ is drawn from $N(0, \sigma^2)$. For each DGP, we consider 12 different scenarios with different number of observations ($n=50, 100, 200, 500$) and the levels of noise ($\sigma=0.5, 1, 2$). Each scenario is replicated 50 times to calculate the in-sample and out-of-sample performances with the mean squared error (MSE) statistic.

Regarding the tuning parameter selection, the optimal hyperparameters (i.e., $C$ and $\varepsilon$) are determined in the SVR and CSVR approaches by using the fivefold cross-validation method. Following \cite{valero-carreras2021support}, the tuning values $C$ and $\varepsilon$ are varied from the multiplier sets $\{0.1, 0.5, 1, 2, 5\}$ and $\{0, 0.001, 0.01, 0.1, 0.2\}$, respectively. For the LCR approach, as in \citeasnoun{mazumder2019a}, we employ the one standard error rule in cross-validation to search the optimal Lipschitz parameter $L$.

In the following experiments, we resort to the pyStoNED package (\citename{Dai2021b}, \citeyear*{Dai2021b}) with the standard solver Mosek (9.2.44). All computations are performed on Aalto University’s high-performance computing cluster Triton with Xeon @2.8 GHz processors, 1 CPU, and 8 GB RAM per task. The simulation code and data are available at the GitHub repository (\url{https://github.com/ds2010/CSVR}).

\subsection{In-sample performance}

We first compare the in-sample performance of CSVR with alternatives. Table \ref{tab: perf1} reports the MSE statistic of each approaches with $n \in \{50,100,200,500\}$, $d \in \{1,2,3\}$, and $\sigma=1$. Note that the MSEs of NCCSVR in the univariate cases are missing due to the fact that the NCCSVR approach requires the multidimensional data space (i.e., $d \ge 2$) (\citename{wang2012multivariate}, \citeyear*{wang2012multivariate}). The optimal hyperparameters in LCR, CSVR, and NCCSVR are prespecified via the standard cross-validation technique. 

Table \ref{tab: perf1} indicates that CSVR exhibits the lowest values of MSE in almost all scenarios. Compared to CSVR, LCR has competitive performance in the univariate case (i.e., DGP I), but its performance deteriorates in multivariate cases.

Compared to the regularized approaches, the traditional CR approach performs poorly in the multivariate cases (i.e., DGP II and III), even though it is actually quite competitive in the univariate setting. After restricting the subgradients in CR, the performance in fitting the true function will be improved as the optimal subgradients cannot take any values for given feasibility. As expected, the performance of each approach deteriorates as more input variables or smaller sample sizes are introduced. 
\begin{table}[H]
\centering
\caption{MSE comparison of different approaches with $\sigma=1$.}
\begin{tabular}{l l l l l l l}
    \hline
    DGP &$n$ &CSVR &SVR &CR &LCR &NCCSVR\\
    \hline
    I &50 & 0.0644 & 0.1099 & 0.0838 & 0.0673 & \multicolumn{1}{c}{--}\\
    &100 & 0.0384 & 0.0694 & 0.0427 & 0.0377  & \multicolumn{1}{c}{--}\\ 
    &200 & 0.0209 & 0.0403 & 0.0252 & 0.0192  & \multicolumn{1}{c}{--}\\
    &500 & 0.0086 & 0.0152 & 0.0107 & 0.0080  & \multicolumn{1}{c}{--}\\
    II &50 & 0.0771 & 0.1125 & 0.2193 & 0.1674 & 0.1085\\ 
    &100 & 0.0386 & 0.0696 & 0.1270 & 0.0904 & 0.0596\\ 
    &200 & 0.0271 & 0.0400 & 0.0829 & 0.0536 & 0.0300\\
    &500 & 0.0083 & 0.0188 & 0.0341 & 0.0222 & 0.0123\\
    III &50 & 0.0758 & 0.0994 & 0.4189 & 0.3198 & 0.1543\\ 
    &100 & 0.0556 & 0.0788 & 0.2765 & 0.2124 & 0.0883\\ 
    &200 & 0.0365 & 0.0484 & 0.1855 & 0.1276 & 0.0453\\
    &500 & 0.0184 & 0.0274 & 0.0974 & 0.0599 & 0.0188\\
    \hline
\end{tabular}
\label{tab: perf1}
\end{table}

To assess the robustness of the CSVR approach, we next consider scenarios with different levels of error variance (i.e., $\sigma$) using a fixed sample size $n=500$ (see Table \ref{tab: perf2}). As expected, we observe that the MSE values increase for all methods as the data noise variance increases. However, the proposed CSVR approach has the smallest MSEs in all cases and is robust to the increasing noise. Note that those methods that use the regularization techniques are more robust to noise than the original CR approach. Overall, CSVR maintains its good performance in different levels of data noise and dimensions.
\begin{table}[H]
\centering
\caption{MSE comparison of different approaches with $n=500$.}
    \begin{tabular}{l l l l l l l}
    \hline
    DGP &$\sigma$ &CSVR &SVR &CR &LCR &NCCSVR\\
    \hline
    I &0.5 & 0.0030 & 0.0043 & 0.0031 & 0.0025 & \multicolumn{1}{c}{--}\\
    &1 & 0.0086 & 0.0152 & 0.0107 & 0.0080 & \multicolumn{1}{c}{--}\\ 
    &2 & 0.0257 & 0.0522 & 0.0392 & 0.0271 & \multicolumn{1}{c}{--}\\
    II &0.5 & 0.0038 & 0.0076 & 0.0090 & 0.0066 & 0.0043\\ 
    &1 & 0.0083 & 0.0188 & 0.0341 & 0.0222 & 0.0123\\ 
    &2 & 0.0214 & 0.0482 & 0.1341 & 0.0755 & 0.0494\\
    III &0.5 & 0.0077 & 0.0114 & 0.0248 & 0.0185 & 0.0070\\ 
    &1 & 0.0184 & 0.0274 & 0.0974 & 0.0599 & 0.0188\\ 
    &2 & 0.0493 & 0.0642 & 0.3874 & 0.1862 & 0.0671\\
    \hline
    \end{tabular}
\label{tab: perf2}
\end{table}

Compared to other regularized approaches, CSVR is relatively more robust when the data noise varies from 0.5 to 2 (see Table \ref{tab: perf2}). While the LCR approach can benefit from the Lipschitz regularization and avoid overfitting, it still uses the squared L$_2$ norm loss function and thus tends to be sensitive to outliers and heteroscedasticity. However, the CSVR approach introduces the $\epsilon$-insensitive loss function inherited from SVR to increase robustness. For smaller $\sigma$ values, as shown in Table \ref{tab: perf2}, it seems that NCCSVR may outperform the proposed CSVR method (i.e. DGP III, $\sigma=0.5$). 

Recall that \cite{wang2012multivariate} have proposed a similar NCCSVR approach, where they introduce convexity/concavity into SVR by using the Hessian matrix and then transform it to a semidefinite programming problem. In contrast, our CSVR approach resorts to the system of Afriat inequalities to ensure the fitted function to be convex/concave. Moreover, the NCCSVR approach can be applied to the multivariate cases only (i.e., $d \ge 2$) and requires one more tuning parameter. To further understand the performance difference between these two similar approaches, we implement the following additional experiment to report the prediction accuracy.

\begin{figure}[H]
	\centering
	\begin{subfigure}[b]{0.475\textwidth}
			\centering
			\includegraphics[width=\textwidth]{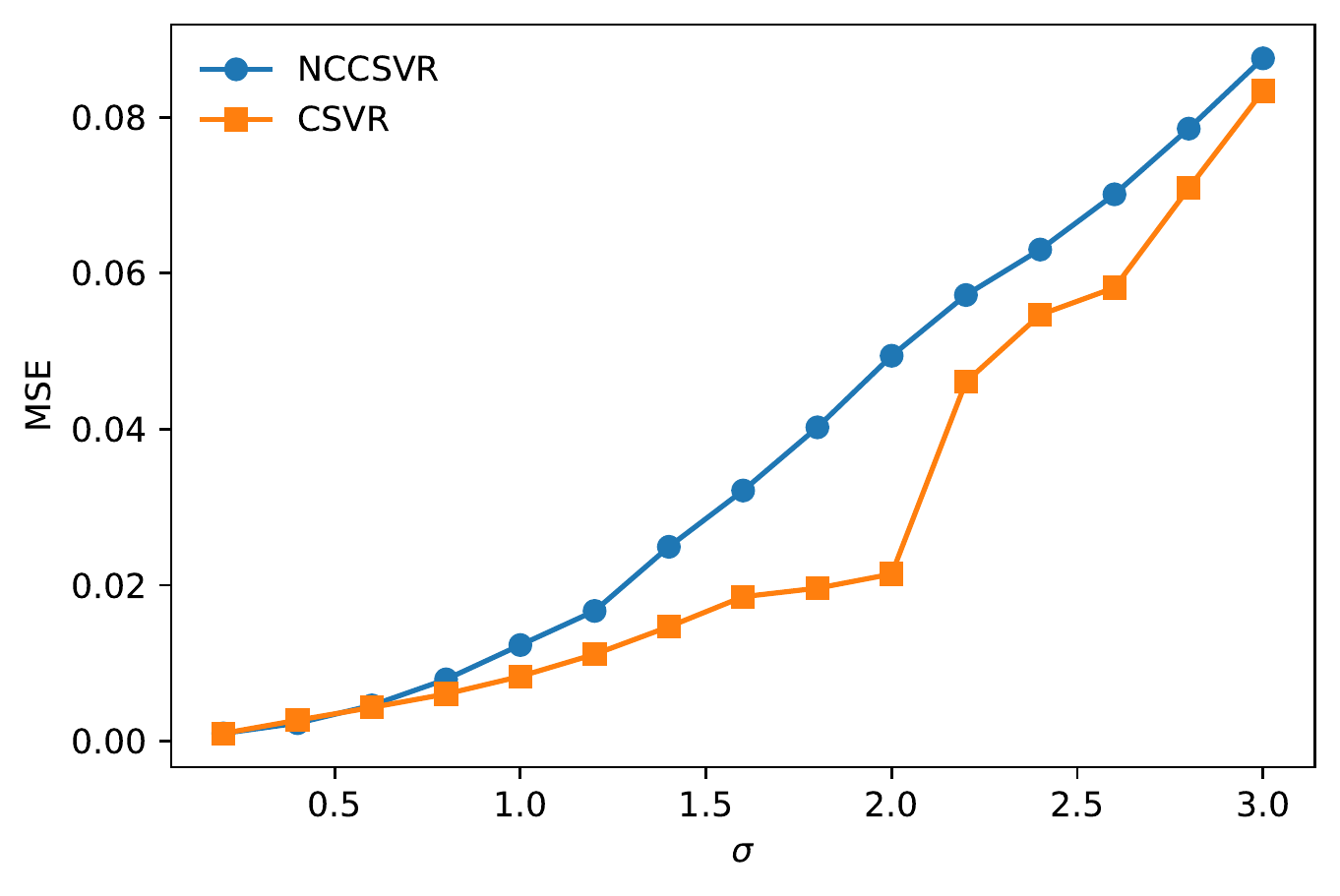}
			\caption[]%
			{{\small $d=2$}}    
	\end{subfigure}
	\hfill
	\begin{subfigure}[b]{0.475\textwidth}  
			\centering 
			\includegraphics[width=\textwidth]{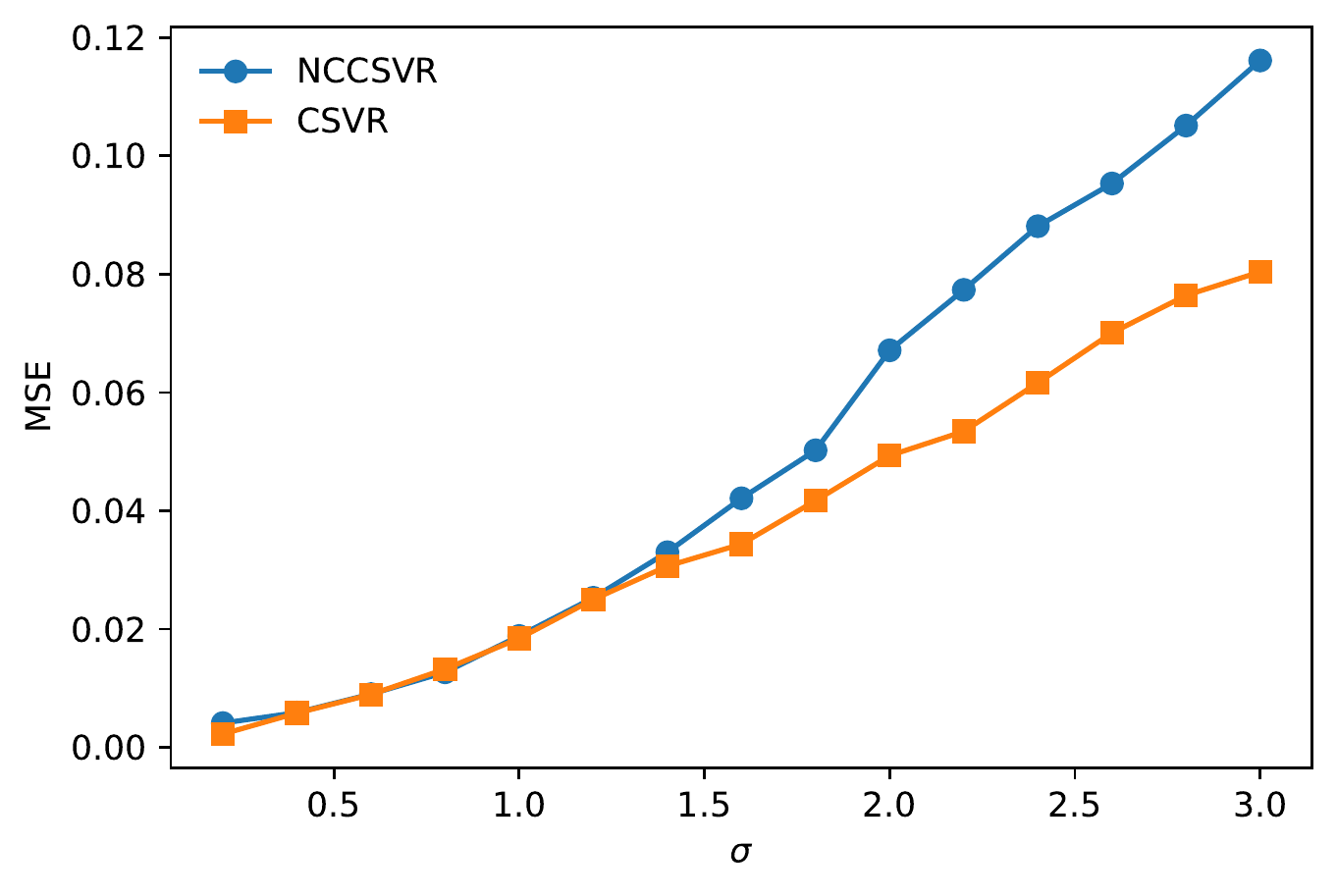}
			\caption[]
			{{\small $d=3$}}   
	\end{subfigure}
	\quad
	\caption[]
	{\small Illustration of the impacts of noise variation on MSE.} 
	\label{fig: sig}
\end{figure}

In previous experiments, we have observed that the variation of the noise $\sigma$ has a significant impact on the MSE values, especially in the cases with small $\sigma$. Here we investigate the impact of noise variation on these two models' performance. Fig.~\ref{fig: sig} depicts the performance of CSVR and NCCSVR when $n=500$, $d=\{2,3\}$, and $\sigma$ varies from 0.2 to 3. Fig.~\ref{fig: sig} a) indicates that the performance of CSVR dominates NCCSVR, whereas Fig.~\ref{fig: sig} b) shows that there are tiny differences in the finite sample performances between NCCSVR and CSVR, but increasing larger differences occur as $\sigma$ grows. This might suggest that imposing the Hessian matrix is not as efficient as the system of Afriat inequalities in terms of overfitting reduction. 
	
We further investigate the impact of outliers. The additional five outliers are drawn from the uniform distribution $U[90, 100]$, and the other normal observations are also drawn from $U [1, 10]$. For the sake of illustration, we simply consider the instances with $n \in \{50, 100\}$, $d \in \{2, 3\}$, and $\sigma = 1$. The in-sample MSEs of each scenario are averaged in a total of 50 replications. As expected, Table \ref{tab: outlier} shows that the CSVR approach performs best among all compared approaches in terms of prediction accuracy. The Lipschitz norm convex regression approach described in \citeasnoun{mazumder2019a} can also control the impact of outliers and, in this case, outperforms the conventional SVR approach, which would have better performance if there were no additional outliers. Furthermore, compared to results reported in Table \ref{tab: perf1}, the results shown in Table \ref{tab: outlier} imply that the additional outliers can lead to worse accuracy for all methods. We also observe that the performance of NCCSVR varies sharply with the parameters, but this does not happen in CSVR. Moreover, the choice of its kernel parameter is highly dependent on the values of input data, which might lead to a limitation in practical applications. Thus it could be hurt for practitioners to decide the range of the three parameters used in NCCSVR. 
\begin{table}[H]
\centering
\caption{MSE comparison of different approaches with additional five outliers.}
    \begin{tabular}{l l l l l l l}
    \hline
    DGP &$n$ &CSVR &SVR &CR &LCR &NCCSVR\\
    \hline
    II & 50 & 0.0979 & 0.1900 & 0.1948 & 0.1477 & 0.1096\\ 
    & 100 & 0.0563 & 0.1175 & 0.1154 & 0.0879 & 0.0595\\ 
    III & 50 & 0.0996 & 0.2251 & 0.4093 & 0.1034 & 0.1556\\ 
    & 100 & 0.0658 & 0.1115 & 0.2074 & 0.0897 & 0.1065\\ 
    \hline
    \end{tabular}
\label{tab: outlier}
\end{table}

\subsection{Out-of-sample performance}

We proceed to investigate out-of-sample performance by considering six scenarios with $d \in \{2, 3\}$, $\sigma\in \{0.5, 1, 2\}$, and $n=500$ for the training set and another 1000 hold-out observations for the test set. We then replicate each scenario 50 times to obtain an empirical distribution of the out-of-sample MSE statistic. Note that the in-sample overfitting may result in low prediction accuracy of the out-of-sample model, and if an estimator is likely to overfit in multidimensional data space, then it will have a smaller in-sample MSE and a larger out-of-sample MSE.

\begin{figure}[H]
    \centering
    \includegraphics[width=0.65\textwidth]{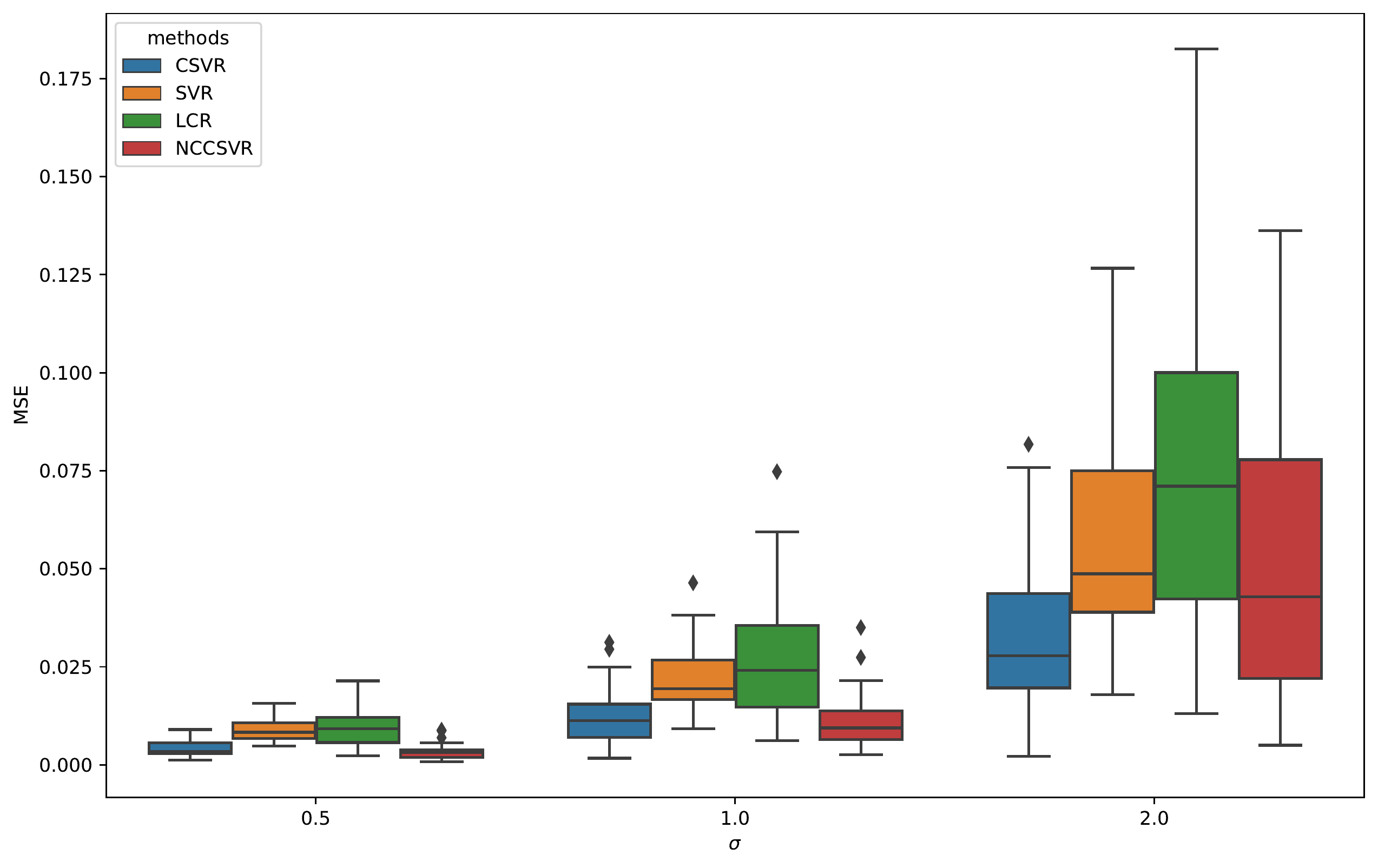}
    \caption{Out-of-sample MSE of the four methods with $d=2$ and $\sigma\in \{0.5,1,2\}$.}
    \label{fig: out1}
\end{figure}

\begin{figure}[H]
    \centering
    \includegraphics[width=0.65\textwidth]{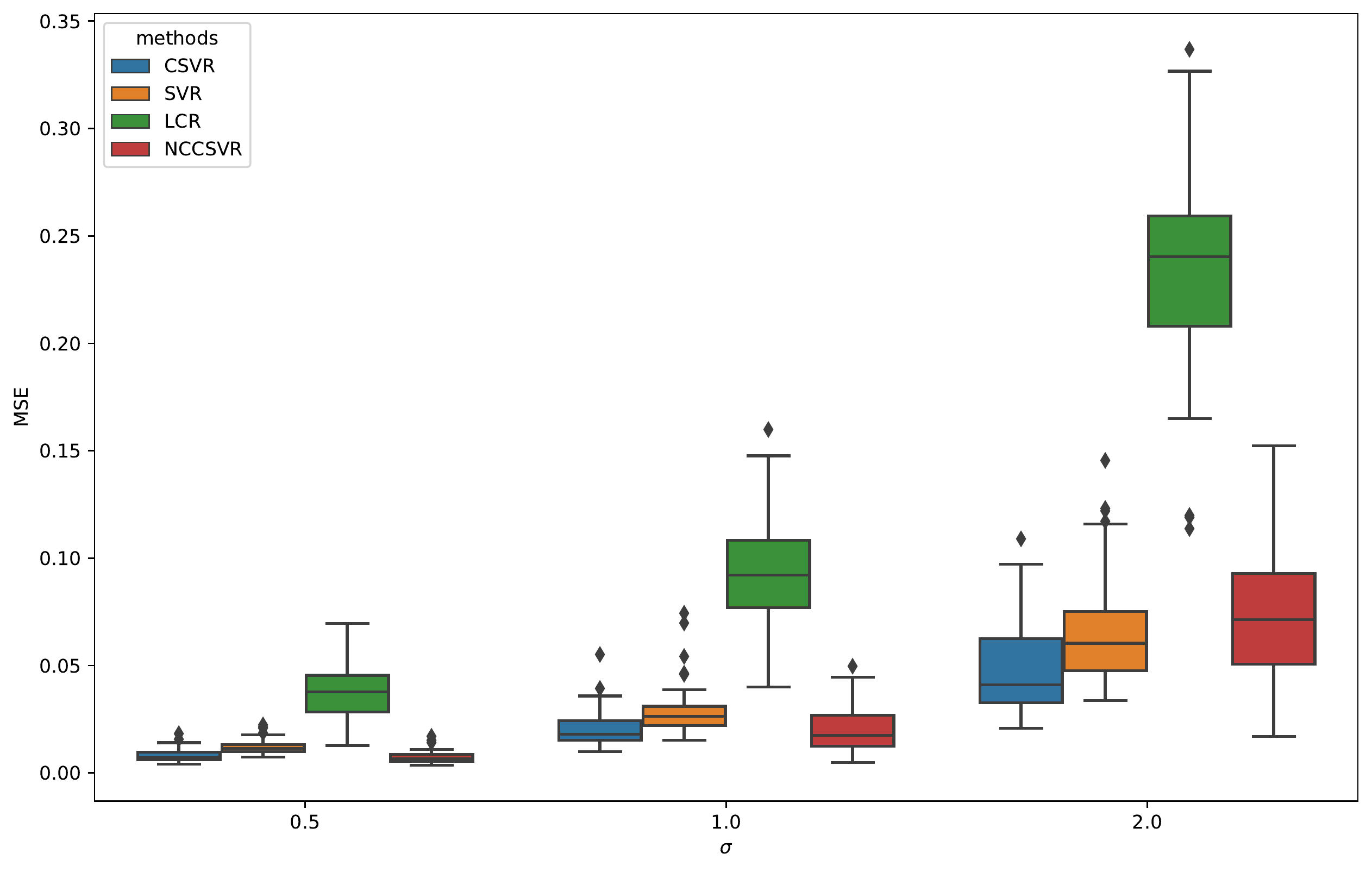} 
    \caption{Out-of-sample MSE of the four methods with $d=3$ and $\sigma\in \{0.5,1,2\}$.}
    \label{fig: out2}
\end{figure}

The boxplots in Figs.~\ref{fig: out1} and \ref{fig: out2} illustrate the distributions of out-of-sample MSE. When comparing traditional CR with regularized alternatives, we observe that CR has a relatively large out-of-sample MSE, which is far more than that of the other regularized approaches. For instance, the values of out-of-sample MSE for CR are $0.121(\pm 0.342)$, $0.454(\pm 1.192)$, and $1.705(\pm 4.300)$ respectively with $d=2$ and $\sigma\in \{0.5, 1, 2\}$. To facilitate a comparison of the most competitive alternatives, we exclude the MSE results for CR from Figs.~\ref{fig: out1} and \ref{fig: out2}.

CSVR performs better than other methods in alleviating the overfitting problem. Compared to the LCR, the three SVR-based approaches seem to perform better in alleviating the overfitting problem. The $\epsilon$-insensitive loss function used in SVR-based approaches is more robust to outliers and large errors than the least squares loss function, which, in turn, helps to build a robust prediction model as demonstrated in Fig.~\ref{fig: out2}. In the experiments, we also find that the out-sample-sample performance of LCR deteriorates rapidly as the data noise increases. 

%
        
\section{Experiments with real data}\label{sec: app}

In this section, we apply the proposed approach to two real-world datasets: the Boston housing data and the NBER-CES manufacturing industry data. Those real datasets have been used extensively in various fields of economics, econometrics, statistics, and machine learning for different purposes such as new models test or algorithms benchmarking (see, e.g., \citename{lin2011vif}, \citeyear*{lin2011vif}; \citename{wang2013estimating}, \citeyear*{wang2013estimating}). In both examples below, all hyperparameters are tuned over from 50 candidate values via the fivefold cross-validation technique, and the in-sample MSE and out-of-sample MSE for each approach are calculated to compare the effectiveness in reducing overfitting. 

\subsection{Boston housing data}

This dataset contains housing price information in the Boston area collected from the StatLib archive.\footnote{
StatLib--Datasets Archive: \url{http://lib.stat.cmu.edu/datasets/boston}.
} 
The data includes 13 variables with 506 observations.\footnote{Note that the dummy variable (i.e., \textit{CHAS}) is excluded from the dataset for the sake of simplicity. Note that CR can handle contextual variables, but this falls beyond the scope of this paper; cf. \citename{johnson2012one} (\citeyear*{johnson2012one}), for further details.} 
Following the commonly used setting, the variable \emph{MEDV} is taken as the response variable of interest, and others are the explanatory variables. The descriptive statistics for all variables are presented in Table \ref{tab: sta}.

Table \ref{tab: beta} reports the descriptive statistics for the estimated coefficients (i.e., $\hat{\alpha}_i$ and $\hat{\boldsymbol{\beta}}_i$) of CSVR, LCR, and CR. We omit SVR and NCCSVR here due to the incomparable dual variable (i.e., $\hat{\boldsymbol{\beta}}_i$) (see \citename{smola2004tutorial}, \citeyear*{smola2004tutorial}). As shown in Table \ref{tab: beta}, all values of $\beta$ for CSVR and LCR lie roughly between -0.5 and 0.5, whereas those values for CR lie in a larger interval, which can be a symptom of overfitting. We also note that although both CSVR and LCR have small $\beta$, LCR obtains coefficients even a bit smaller than CSVR. Furthermore, LCR produces a more flat function $f$, but that does not mean a better performance automatically. 
 
To evaluate the prediction accuracy, we divide the 506 samples randomly into 405 training samples and 101 test samples and then compute the in-sample and out-of-sample prediction error in terms of MSE. The calculated in-sample and out-of-sample MSEs and estimated standard deviations are demonstrated in Table \ref{tab: boston}. Overall, as expected, we observe that the CSVR method achieves the best prediction performance. The results show that the restriction on the subgradients or regularization described in Section \ref{sec: csvr} leads to a flatter estimated function that can overcome overfitting. Although LCR is also developed for solving overfitting problems, our approach still outperforms the LCR method. A possible explanation is that owing to the structure of the SVR-based approaches, CSVR can achieve a good bias-variance trade-off. Compared to the non-regularized method (i.e., CR in Table \ref{tab: boston}), all methods benefit from additional regularization in terms of the out-of-sample prediction accuracy. However, note that CR yields the lowest in-sample MSE in this empirical data.
\begin{table}[H]
\centering
\caption{Descriptive statistics for estimates of all explanatory variables.}
    \begin{tabular}{lrrrrrr}
    \hline
    \multirow{2}{*}{} &\multicolumn{2}{c}{CSVR} &\multicolumn{2}{c}{LCR} &\multicolumn{2}{c}{CR}  \\ 
    \cline{2-3} \cline{4-5} \cline{6-7}
    &\multicolumn{1}{c}{Mean} &\multicolumn{1}{c}{Std.Dev.} 
    &\multicolumn{1}{c}{Mean} &\multicolumn{1}{c}{Std.Dev.} 
    &\multicolumn{1}{c}{Mean} &\multicolumn{1}{c}{Std.Dev.} \\
    \hline
    $\hat{\beta}_{CRIM}$    & 0.08    &   0.17    & -0.01     &   0.08  &   2.66 &   10.47 \\
    $\hat{\beta}_{ZN}$      & 0.12    &   0.15    & 0.07      &   0.06  &   0.80  &   9.22 \\
    $\hat{\beta}_{INDUS}$   & -0.02   &   0.22    & -0.03     &	  0.11  &	2.59 &	12.29 \\
    $\hat{\beta}_{NOX}$     & 0.00   &	0.04    &	0.00     &	0.02  &	147.61 &	963.24 \\ 
    $\hat{\beta}_{RM}$      & 0.19    &	0.34    &	0.05     &	0.10  &	11.10  &	64.63 \\
    $\hat{\beta}_{AGE}$     & -0.02   &	0.19    &	-0.01     &	0.08  &	-0.39   &	2.87 \\ 
    $\hat{\beta}_{DIS}$     & -0.17   &	0.34    &	-0.04     &	0.10  &	-0.60   &	48.50 \\ 
    $\hat{\beta}_{RAD}$     & 0.09    &	0.26    &	0.06     &	0.10  & -5.01   &	20.08 \\
    $\hat{\beta}_{TAX}$     & -0.01   &	0.05    &	-0.01     &	0.02  &	0.09    &	0.85 \\
    $\hat{\beta}_{PTRATIO}$ & -0.18   &	0.33    &	-0.06     &	0.11  &	1.89    &	42.85 \\
    $\hat{\beta}_{B}$       & -0.21   &	0.26    &	-0.08     &	0.10  &	-0.25   &	0.88 \\
    $\hat{\beta}_{LSTAT}$   & -0.42   &	0.29    &	-0.28     &	0.12  &	0.24    &	7.68 \\
    $\hat{\alpha}$          & 132.61  &	103.68   &	69.44    &  39.31 &	276.42  &  1264.22 \\
    \hline
    \end{tabular}
\label{tab: beta}
\end{table}

\begin{table}[H]
\centering
\caption{In-sample and out-of-sample MSEs: Boston housing data.}
\begin{threeparttable}
    \begin{tabular}{l l l l l l l}
    \hline
    MSE &CSVR &SVR &CR &LCR &NCCSVR\\
    \hline
    Out-of-sample MSE &38.44(5.04) & 64.00(14.28) & 2334.17(933.04) & 43.72(6.47) &42.72(9.28)\\
    In-sample MSE &21.43(0.86) & 62.57(2.90) & 1.41(0.33) & 35.05(1.52) &40.17(2.05)\\ 
    \hline
    \end{tabular}
\begin{tablenotes}
\footnotesize
    \item[] \textit{Note}: standard deviation in parentheses. 
\end{tablenotes}
\end{threeparttable}
\label{tab: boston}
\end{table}

\subsection{NBER-CES manufacturing industry data}

The data consist of 473 manufacturing industries (the 1997 6-digit NAICS codes) and are collected from the NBER-CES Manufacturing Industry Database.\footnote{NBER database: \url{https://www.nber.org/research/data/nber-ces-manufacturing-industry-database}.
} 
In this application, we apply the present approaches to estimate the production function, where, as in \citeasnoun{wang2012multivariate}, the input includes the capital (INVEST), labor (PAY), and raw materials (MATCOST), and the output is the value added (VADD). Furthermore, the training and test set are 376 and 97 observations, respectively. 

CSVR's in-sample and out-of-sample MSEs are $0.53 \times 10^7$ and $0.99 \times 10^7$ (see Table \ref{tab: nber}). While the in-sample MSE value of CSVR is not the lowest in comparison with other shape-constrained approaches, the out-of-sample MSE remains the lowest. This is perhaps due to the fact that there is a good trade-off between model performance and overfitting alleviation. 
\begin{table}[H]
\centering
\caption{In-sample and out-of-sample MSEs ($\times 10^7$): NBER-CES manufacturing industry data.}
\begin{threeparttable}
    \begin{tabular}{l l l l l l l}
    \hline
    MSE &CSVR &SVR &CR &LCR &NCCSVR\\
    \hline
    Out-of-sample MSE &0.99 & 3.96 & 74.78 & 1.09 &1.83\\
    In-sample MSE &0.53 & 3.94 & 0.37 & 0.91 &1.01\\ 
    \hline
    \end{tabular}
\end{threeparttable}
\label{tab: nber}
\end{table}

It is worth noting that CR has the best in-sample fit in the applications but not in the MC simulations. It is because the MSE is measured differently. In the simulations, we measure the deviation between the estimated function $f$ and the true function $F$. In the applications, the true function $f$ is unknown, and we thus measure the deviation of the predictions from the observed $y$. This is not the same MSE because $y = f(x) + \epsilon$ also includes the noise. CR will always minimize the MSE wrt y, but due to overfitting not wrt $f(x)$. 

In conclusion, both simulations and real-world applications demonstrate that all regularized shape-constrained methods have a superior ability to control overfitting, but CSVR would be more appealing than other regularized shape-constrained methods because of its simplicity, capacity for univariate regression, and robust performance.

%

\section{Conclusions}\label{sec: con}

Overfitting is a commonly seen phenomenon in nonparametric regression. To mitigate the effects of overfitting, we have introduced a new approach called convex support vector regression, which effectively combines the key elements of support vector regression and convex regression. The paper investigates the finite sample performance of the developed CSVR approach in contrast to other state-of-the-art regression methods through Monte Carlo simulations. Additional two real-world datasets are also used to test and compare the performance of these approaches. We hope that the proposed approach can help to further bridge the gaps between the data-driven estimation approaches known in econometrics and statistics, machine learning, and operations research and management science. 

The evidence from the simulations indicates that CSVR performs at least as well as LCR and much better than SVR and traditional convex regression. Two real-world applications also show that our approach outperforms other state-of-the-art regression methods. The regularized convex regression model can help to alleviate the overfitting problem, also owing to its insensitive loss function and robustness in the presence of outliers. 

In this paper, we have restricted attention to regularizations known in the literature, but there could be more efficient ways to restrict the domain of subgradients (e.g., weight-restricted regression). Another promising future research direction is developing the simple and fast algorithmic framework of convex support vector regression. The main bottleneck of CSVR is that the full problem (\ref{csvr}) has $n(n-1)$ constraints and thus becomes computationally inefficient for more than a few thousand observations. Furthermore, we have deliberately kept away from statistical inferences, and further work in this direction, e.g., exploring the asymptotic property of CSVR, would be needed. 

%

\section*{Acknowledgments}

We acknowledge the computational resources provided by the Aalto Science-IT project. Zhiqiang Liao gratefully acknowledges financial support from the Foundation for Economic Education (Liikesivistysrahasto) [grant no. 210038]. Sheng Dai gratefully acknowledges financial support from the Foundation for Economic Education (Liikesivistysrahasto) [grants no. 180019, 190073, 210075] and the HSE Support Foundation [grant no. 11--2290]. 

%

\baselineskip 12pt
\bibliography{References}

%

\newpage
\section*{Appendix}\label{sec:app}

\renewcommand{\thetable}{A\arabic{table}}
\setcounter{table}{0}
\renewcommand{\thetable}{A\arabic{table}} 

\begin{table}[H]
\centering
\caption{Descriptive statistics of all variables: the Boston housing data.}
    \setlength\tabcolsep{2pt}
    \begin{tabular}{llrrrr}
    \hline
    Variable &Description &Mean &Std. Dev. &Min. &Max.\\
    \hline
    CRIM & per capita crime rate by town & 3.61 & 8.60 &0.01 &88.98\\
    ZN & proportion of residential land zoned & 11.36 & 23.32 &0.00 &100.00\\ 
    INDUS & proportion of non-retail business acres per town & 11.14 & 6.86 &0.46 &27.74\\
    NOX & nitric oxides concentration & 0.55 & 0.11 &0.36 &0.87\\ 
    RM & average number of rooms per dwelling & 6.28 &  0.70 &3.56 &8.78\\
    AGE & proportion of owner built prior to 1940 & 68.57 &  28.15 &2.90 &100.00\\ 
    DIS & weighted distances to city centres & 3.80 & 2.11 &1.13 &12.13\\ 
    RAD & index of accessibility to radial highways & 9.55 & 8.71 &1.00 &24.00\\
    TAX & full-value property-tax rate per \$10,000 & 408.24 & 168.54 &187.00 &711.00\\
    PTRATIO & pupil-teacher ratio by town & 18.46 &  2.16 &12.60 &22.00\\
    B & black proportion of population & 356.67 &  91.29 &0.32 &396.90\\
    LSTAT & proportion of population that is lower status & 12.65 & 7.14 &1.73 &37.97\\
    MEDV & median value of owner-occupied homes& 22.53 & 9.20 &5.00 &50.00\\
    \hline
    \end{tabular}
\label{tab: sta}
\end{table}

\end{document}